# Will Southeast Asia be the next global manufacturing hub? A multiway cointegration, causality, and dynamic connectedness analyses on factors influencing offshore decisions

Haibo Wang
Department of International Business and Technology Studies, Texas A&M International University, Laredo, Texas, USA
Lutfu S. Sua
Department of Management and Marketing, Southern University and A&M College, Baton Rouge, LA, USA
Jun Huang
Department of Management and Marketing, Angelo State University, San Angelo, TX, USA
Jaime Ortiz
Department of International Business and Entrepreneurship, The University of Texas Rio Grande Valley, Edinburg, TX 78539, USA
Bahram Alidaee
Department of Marketing, School of Business Administration, University of Mississippi, Oxford, MS, USA

**Abstract**

The COVID-19 pandemic has compelled multinational corporations to diversify their global supply chain risk and to relocate their factories to Southeast Asian countries beyond China. Such recent phenomena provide a good opportunity to understand the factors that influenced offshore decisions in the last two decades. We propose a new conceptual framework based on econometric approaches to examine the relationships between these factors. Firstly, the Vector Auto Regression (VAR) for multi-way cointegration analysis by a Johansen test as well as the embedding Granger causality analysis to examine offshore decisions--innovation, technology readiness, infrastructure, foreign direct investment (FDI), and intermediate imports. Secondly, a Quantile Vector Autoregressive (QVAR) model is used to assess the dynamic connectedness among Southeast Asian countries based on the offshore factors. This study explores a system-wide experiment to evaluate the spillover effects of offshore decisions. It reports a comprehensive analysis using time-series data collected from the World Bank. The results of the cointegration, causality, and dynamic connectedness analyses show that a subset of Southeast Asian countries have spillover effects on each other. These countries present a multi-way cointegration and dynamic connectedness relationship. The study contributes to policymaking by providing a data-driven innovative approach through a new conceptual framework.

1. **Introduction**

The disruption of the supply chain during the global COVID-19 pandemic (GCP) has forced many multinational corporations to reassess their supply chain risks. For example, Apple is considering moving its iPhone production line out of China[1]. Supply chain disruptions, compounded by the Russia-Ukraine conflict, entice multinational corporations to make structural changes to their supply chains. Recent studies indicate that a vast majority of companies aim to increase the flexibility, agility, and resiliency of their supply chains[2]. Although the main focus has been on the localization of supplier networks, a longer-term goal is the diversification of the supply chain to address barriers such as insufficient local suppliers, high investment requirements, and cost disadvantages. Despite these challenges, developing an integrated supply chain, diversifying the supplier base, and forming more regionally concentrated value chains, especially in Asia and Europe, remains a priority for the sustainable competitiveness of multinational corporations[3].

In the past three decades, China has become the world's top manufacturing hub.[4]. However, the US-China trade war has shifted the manufacturing hub into five new locations: Vietnam, Mexico, India, Malaysia, and Singapore[5]. Three of these global hubs are in Southeast Asia. The region was transitioning to become a manufacturing hub until 1997 when the Asian financial crisis[6] (AFC) halted the growth and the spread of industrialization in the region. During 1996-1997, the AFC caused the closure of 16 banks in Indonesia. Similarly, the Global Financial Crisis (GFC) in 2008-2009 triggered a downturn in economic development due to an export slowdown. While countries in the region were slowly recovering from GFC and picking up fast growth, the GCP disrupted the global supply chains on a much larger scale. In this study, we examine the spillover effects of offshore decision factors on Southeast Asian countries, and we discuss the dynamic connectedness among these countries.

The GCP has forced multinational corporations to diversify their global supply chain risk and to relocate their factories to Southeast Asian countries beyond China. However, the relocation and redesign of the global supply chain cannot be completed overnight. The readiness of a newly established supply chain depends on careful long-term plans for infrastructure, trade and finance, and the workforce. To examine the cointegration and causality effects of factors on offshore decisions--innovation, technology readiness, infrastructure, foreign direct investment (FDI), and intermediate imports we propose the following research questions (RQ):

RQ1: will countries that invested in infrastructure/technology receive more Net FDI inflows?

---

[1] https://www.wsj.com/articles/apple-china-factory-protests-foxconn-manufacturing-production-supply-chain-11670023099
[2] https://www.mckinsey.com/capabilities/operations/our-insights/how-covid-19-is-reshaping-supply-chains
[3] https://www.mckinsey.com/featured-insights/innovation-and-growth/globalization-in-transition-the-future-of-trade-and-value-chains
[4] https://globalupside.com/top-10-manufacturing-countries-in-the-world/
[5] https://internationalfinance.com/top-5-global-manufacturing-hubs-making/
[6] https://www.federalreservehistory.org/essays/asian-financial-crisis

RQ2: will countries that invested in infrastructure/technology experience improvements in innovation?

RQ3: will countries that receive more net inflows of FDI bring more manufacturing outputs such as intermediate imports, which are the outcomes of the offshoring decision?

RQ4: What is the spillover effect on offshore decision factors between China and Southeast Asia countries?

RQ5: what is the dynamic connectedness between China and Southeast Asia countries in the process of offshoring shift?

This paper makes the following novel contributions. First, we propose a new conceptual framework of multi-way co-integration, causality, and dynamic connectedness analysis on time-series data from multiple countries with key variables of economic development and offshoring decisions. Second, we implement a standard vector autoregressive (VAR) estimation model for predicting future manufacturing shifts and identifying the variables that are important for those shifts. Third, a quantile vector autoregressive (QVAR) model examines offshore decisions and reveals the dynamic interconnectedness among countries during the shifts.

The rest of the paper is organized as follows. Section 2 provides a comprehensive review of offshore decisions using co-integration, causality, and dynamic connectedness analyses. A new conceptual framework of multi-way cointegration, causality, and dynamic connectedness analyses is presented in Section 3, followed by a report of the results and findings in Section 4. We summarize the implications for theory and practice in Section 5, and we finally provide conclusions in Section 6.

## 2. Literature Review
### 2.1. Offshoring

The cross-border relocation of operations has become a common practice for many businesses to remain competitive [1], coupled with advanced information and communication technology (ICT), transportation technology and infrastructure, and globalization dynamics.

The offshoring-competitiveness relationship has been studied both at the firm level [2-4] as well as the national level [5-7]. Drivers of offshoring decisions [8, 9], the impact of offshoring on competitiveness through FDI [10, 11], and location decisions [12] are among the topics investigated. Prime et al. (2012) focused on the relationship between competitiveness and offshoring decisions to report that offshoring advanced operations is a strategic move to achieve international competitiveness and to tap into foreign knowledge, while less advanced operations are offshored mainly to take advantage of lower costs [13].

Although there are contradictory arguments about the relationship between offshoring and innovation [14, 15], many firms choose to offshore their routine tasks enabling them to focus more on cognitive tasks [16].

Amendolagine (2014) reports that offshoring firms are larger, more innovative, and have a higher capital/labor ratio [17]. While there are many studies on the relationship between innovation and competitiveness [18, 19], it is difficult to predict the magnitude of the innovation on the competitiveness of an economy mainly due to the complex data and methodology requirements [20].

FDI is an investment category that represents the interest and influence of a business entity established in one country over another business entity in a different country [21]. Motivated mainly by differences in the cost of operations, FDI is influenced by factors such as cost and productivity of labor, market size, incentives, and regulations. Thus, its impact on growth [22] and competitiveness [23] depends on a long list of factors [24-33]. The region-specific nature of FDI has led to a wide range of studies on FDI in the Asia-Pacific region [34], Europe [35-37], Africa [38], and Latin America [39]. Basu et al. (2003) show a bi-directional causality relationship between FDI and GDP growth in developing countries [40]. They suggested that FDI can contribute to economic growth through productivity increases resulting from spillovers to local firms [41].

Offshoring is defined as the relocating of a part of the production process overseas and can be in several forms, such as FDI and international outsourcing [42]. International outsourcing entails the transfer of certain business operations, either partially or entirely, from one nation to another[7]. For instance, western-based OEM may opt to outsource the manufacturing of its products to a supplier located in a country with lower production costs, thereby leveraging reduced wages. In this scenario, the supplier typically receives intermediate goods, assembles them, and subsequently exports the final products back to the originating company. Consequently, apart from Foreign Direct Investment (FDI), we also utilize the importation of intermediate goods from Southeast Asian countries as an indicator to gauge the extent of offshoring within these nations [43, 44].

### 2.2 Cointegration

Dutta and Roy (2005) applied a dynamic causal model to determine the mechanics by which the factors used in the related literature interact to affect offshoring decisions [45]. Usman and Bashir (2022) used Granger causality to investigate the causality between intermediate imports and economic growth measured by GDP per capita and reported a bi-directional causal relationship [46]. Feng et al. (2023) constructed time-varying tail risk networks to investigate systemic risk spillovers in the Belt and Road stock markets during the period 2008 - 2021 [47]. Diebold and Yilmaz (2009) proposed a framework where a volatility spillover measure based on forecast error variance decompositions from VARs was introduced [48]. Such a framework was further improved by including directional volatility spillovers to uncover the system of connectedness under static and time-varying settings [49].

---

[7] https://usitc.gov/research_and_analysis/trade_shifts_2017/specialtopic.htm

To examine the relationships between the factors influencing offshore decisions, one of the econometric approaches utilized in this study is Vector Auto Regression (VAR) for multi-way cointegration analysis.

### 2.3 Dynamic connectedness, VAR, and QVAR models

Investigating dynamic network spillovers and the potential impacts of shocks on the relevant networks has increased its popularity over the years, resulting in various extensions of VAR models [50]. Quantile VAR (QVAR) involves modeling each quantile of the distribution to measure the dynamic connectedness in a network.

Tiwari et al. (2022) utilized a rolling window-based QVAR when investigating the conditional volatility spillover before and during COVID-19 [51]. Lin and Su (2021) applied the dynamic connectedness index method based on the Time-Varying Parameter Vector Autoregression (TVP-VAR) model as the main method to study how the spillover relationships change over a time horizon while avoiding the defects of a rolling window estimation [52]. Antonakakis et al. (2019) applied an extended TVP-VAR connectedness approach to investigate uncertainty spillovers between Greece and Europe through the GFC [53]. Building upon the work of Diebold and Yilmaz [49], Gabauer et al. (2023) proposed model-free or unconditional connectedness measures, investigating only the bivariate relations, thus greatly increasing the network size [54].

While there is abundant research that focuses on only one country or uses simple regression models with few variables, works are reporting a system-wide experiment on the region using VAR-QVAR to analyze the time-series data in terms of dynamic connectedness and cointegration.

### 3. Research Methodology

This study concerns the time-series data, and the workflow of the research method is given in Figure 1.

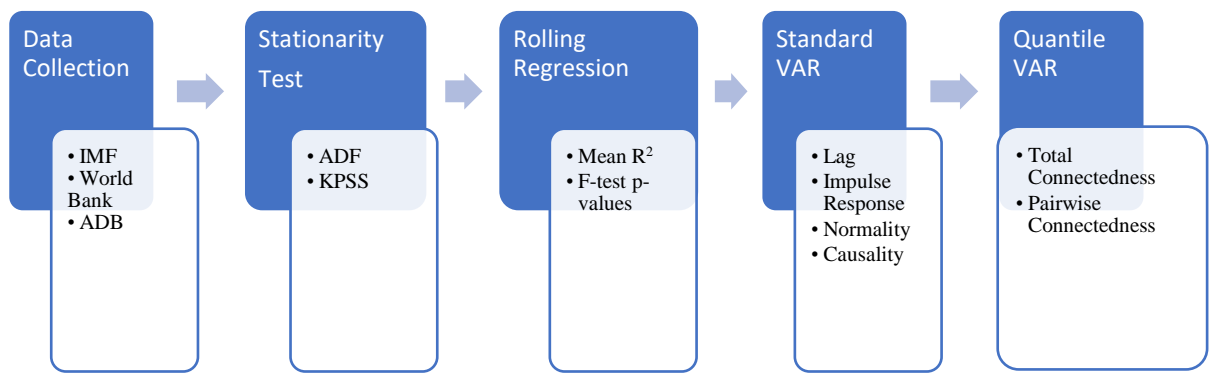

Figure 1. The workflow of time-series data analysis

The Augmented Dickey-Fuller (ADF) and Kwiatkowski-Phillips-Schmidt-Schin (KPSS) tests are used to confirm the stationarity of the time series. As opposed to the ADF test, the null hypothesis in KPSS, which is a linear regression-based test, is that the time series is stationary which means lower *p*-values than the designated significance level indicates non-stationary series. Both tests are applied to confirm stationarity. Occasional differences in the test results indicate that the series are trend stationary as opposed to strict stationary.

To answer research questions RQ1-RQ3, a rolling regression analysis is deployed. Rolling regressions are one of the simplest models for analyzing changing relationships among variables over time. They use linear regression but allow the data set to change over time. In most linear regression models, parameters are assumed to be time-invariant and thus should not change over time. Rolling regressions estimate model parameters using a fixed window of time over the entire data set. They are implemented by the Rolling Ordinary Least-Squares (OLS) function.

The standard OLS formula is given as:

$$Min\ S = \sum_{i=1}^{n}(\hat{\epsilon}_i)^2 = \sum_{i=1}^{n}(y_i - \hat{y}_i)^2 = \sum_{i=1}^{n}(y_i - b_i x_i - b_0)^2 \qquad (1)$$

Where, $\hat{y}_i$ is the predicted value for the *i*th observation, $y_i$ is the actual value for the *i*th observation, $\hat{\epsilon}_i$ is error/residual for the *i*th observation, $n$ is the number of observations. To compute the values of $b_i$ and $b_0$, we can take a partial derivative for each coefficient and equate it to zero. There are several assumptions on data properties involved in testing rolling regression on time series data. First, the dependent and independent variables need to be stationary and the statistical properties of a sample of each time interval present the average statistical properties of the data within the time interval. Since we slide the window across the dataset instead of sampling from the whole dataset as well as taking a log of the data used for calculation, this assumption is satisfied. If the dependent and independent variables are co-integrated, then the results of a rolling regression are valid even if the data are not stationary. The second assumption is that the mean value of the residuals of the estimated rolling OLS model is equal to zero and there is no perfect multicollinearity. This assumption is applied to any general regression model.

The rolling regression model for RQ1 is given:

$$fdi_t = \alpha_t^1 + \beta_t^1 * trd_t + \gamma_t^1 * ifr_t + \epsilon_t^1 \qquad (2)$$

The rolling regression model for RQ2 is given:

$$inv_t = \alpha_t^2 + \beta_t^2 * trd_t + \gamma_t^2 * ifr_t + \epsilon_t^2 \qquad (3)$$

The rolling regression model for RQ3 is given:

$$iep_t = \alpha_t^3 + \beta_t^3 * fdi_t + \epsilon_t^3 \qquad (4)$$

To answer RQ1, RQ, RQ3, RQ4, and RQ5, this study proposes a new conceptual framework for multi-way cointegration, causality, and dynamic connectedness analyses to examine relationships of factors on offshore decisions--innovation, Technology-readiness, infrastructure, foreign direct investment (FDI), intermediate imports. Figure 2 illustrates the new conceptual framework developed:

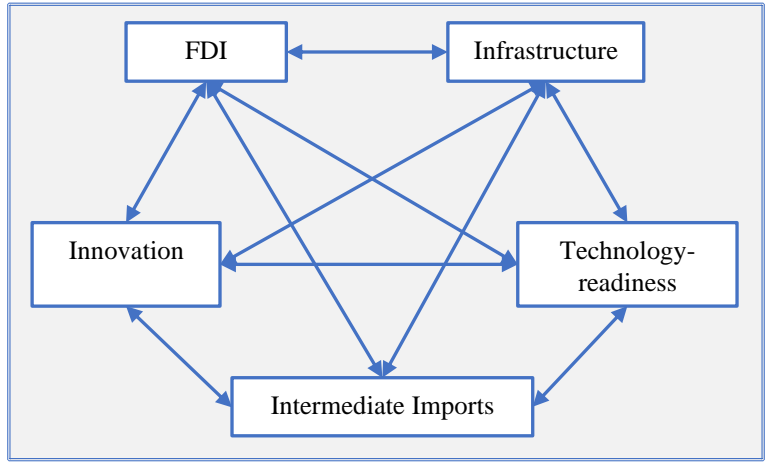

**Figure 2.** Conceptual framework

VAR in level and VAR in difference models for cointegration and causality analysis are deployed to investigate the integration relationship among multiple variables. As a statistical model, VAR captures the relationship among multiple variables over a period. The VAR model can be expressed as below:

$$x_t = \alpha + \emptyset_1 x_{t-1} + \ldots + \emptyset_n x_{t-n} + w_t \tag{5}$$
$$w_t \sim Normal(0, \sum_w)$$

Causality and cointegration are tested before and after each critical event. The estimation models of VAR have two types: VAR in levels and VAR in differences. The basic model is "VAR in level" with the assumption of stationary time series. We have two or four variables so we cannot have stationary on all of them at the same time. The safest bet is to test both models to see if the lags are different. "VAR in differences" can cover the scenario when one variable is stationary, and the other is not. In this study, the Johansen test is used to test the cointegration relationship [55].

The Granger causality test is used to test the causality between the variables under investigation. Its null hypothesis states that one variable does not explain the variation in another. Probability values less than a pre-specified level led to the rejection of the null hypothesis, thus concluding in causality between the two variables. The Granger-causality model not only calculates the precision of the prediction of the variable of interest based on its past values but also calculates the precision of the prediction of the variable of interest based on the moving average of past values of a second variable. To evaluate the goodness of fit of the Granger-causality model, the root mean square (RMS) is calculated under the assumption that the residuals of the time series are random and normally distributed.

$$RMS = \sqrt{\frac{\sum_{i=1}^{N-L}(\check{x}_i - x_i')^2}{N-L}} \quad (6)$$

In (3), *N* is the number of data points in the time series, *L* is the order of the model, and *N-L* is the number of data points to construct the model. $\check{x}$ is the vector for the data points that were subtracted from the original data by the first *L* data.

To determine the value of lag differenced or augmented terms, the Akaike Information Criterion (AIC), Hannan Quinn (HQ), Schwarz Criterion (SC), and Final Prediction Error (FPE) criteria are used. In most cases, all four criteria have the same value for lag selection. If the same lag is indicated by these criteria, it should be chosen and over-parametrization that would result from an additional lag should be avoided. When there are differences among the four criteria, the lag value of SC is selected because the smaller number indicates the fitness of the lag and is suitable for VAR models.

Cointegration between multiple variables is tested using the Jarque-Bera (JB) goodness-of-fit test to evaluate the distribution of residuals of the VAR model for the variables. The test provides skewness and kurtosis computations to determine normality. The residuals must be randomly distributed for normality. *P*-values <0.05 indicate the validity of the VAR model, thus a bidirectional cointegration. Pairwise cointegration between two variables implies the existence of system-wide cointegration.

$$\begin{bmatrix} fdi_t \\ trd_t \\ inv_t \\ ifr_t \\ iep_t \end{bmatrix} = A_0 + A_1 \begin{bmatrix} fdi_{t-1} \\ trd_{t-1} \\ inv_{t-1} \\ ifr_{t-1} \\ iep_{t-1} \end{bmatrix} + A_2 \begin{bmatrix} fdi_{t-2} \\ trd_{t-2} \\ inv_{t-2} \\ ifr_{t-2} \\ iep_{t-2} \end{bmatrix} + \cdots + A_p \begin{bmatrix} fdi_{t-p} \\ trd_{t-p} \\ inv_{t-p} \\ ifr_{t-p} \\ iep_{t-p} \end{bmatrix} + CD_t + u_t \quad (7)$$

*where:*

$fdi_t$: FDI
$inv_t$: Innovation
$trd_t$: Technology-readiness
$ifr_t$: Infrastructure
$iep_t$: Intermediate imports

To assess the connectedness among Southeast Asian countries for RQ5, we implement QVAR for dynamic connectedness analysis. Introduced by [56], QVAR cast the standard VAR models with a quantile regression model to measure the dynamics of the selected quantiles. The QVAR model has advantages over the standard rolling window analysis such as rolling regression, and standard VAR models with constant parameters. Unlike the rolling regression model that sets the rolling window size arbitrarily, QVAR defines the multiple quantiles instead of arbitrary rolling window size to avoid the loss of information in computing dynamic connectedness. QVAR is less sensitive to outliers using a recursive information set than the standard VAR with constant parameters.

In QVAR, a recursive information set $\Omega$ is used to model multiple quantiles [56]. For the lagged values of $\tilde{Y}_t$ and the contemporaneous value of $\tilde{Y}_{1,t+1}$, $\Omega_{1t} = \{\tilde{Y}_t, \tilde{Y}_{t-1}, \cdots\}$ and $\Omega_{it} = \{\tilde{Y}_{i-1,t+1}, \Omega_{i-1,t}\}$.

For $p$ multiple distinct quantiles, $0 < \theta_1 < \theta_2 < \cdots < \theta_p < 1$, the QVAR model is given:

$$Y_{t+1} = \omega + A_0 Y_{t+1} + A_1 Y_{t+1} + \epsilon_{t+1}, \quad P(\epsilon_{i,t+1}^{\theta_j} < 0|\Omega_{1t}) = \theta_j, \; P = 1,\cdots,n, j = 1,\cdots,p, \quad (8)$$

The error term $\epsilon_{t+1}$ is quantile specific and satisfies the condition of $P(\epsilon_{i,t+1}^{\theta_j} < 0|\Omega_{1t}) = \theta_j$. $Y_t = 1_p \otimes \tilde{Y}_t$, where $1_p$ is a $p$-vector of ones. The matrices $A_0 = I_p \otimes \widetilde{A_0}$ with lower triangular matrices with zeros in the main diagonal and $A_1 = I_p \otimes \widetilde{A_1}$, to avoid trivial multicollinearity problems.

The forecast model of QVAR can be described as branches of a tree. For $p$ multiple distinct quantiles, the starting node, $\tilde{Y}_{1,t+1}$, has $p$ branches (the $p$ quantiles). At the end of each branch, there are $p$ more branches for $\tilde{Y}_{2,t+1}$, and so on. The general form of QVAR is given:

$$\bar{Y}_{t+1} = \bar{\omega} + A^0 \bar{Y}_{t+1} + A^1 \bar{Y}_t + \epsilon_{t+1}, \tag{9}$$

where $A^0 = \begin{bmatrix} A_0 & 0 & 0 \cdots & 0 \\ 0 & & & \vdots \\ \vdots & & \ddots & \\ 0 & \cdots & & 0 \end{bmatrix}$ and $A^1 = \begin{bmatrix} A_1 & A_2 & \cdots & A_q \\ I_{np} & 0 & \cdots & 0 \\ \vdots & \vdots & \ddots & \vdots \\ 0 & \cdots & I_{np} & 0 \end{bmatrix}$

QVAR model provides the natural environment to measure impulse response to a given shock by defining a set of future tail quantiles of interest and to predict the outcomes of variables conditional on the chosen shock. The structural quantile impulse response function is given in terms of structural shocks:

$$Y_t = (I_{np} - A_0)^{-1}\omega + (I_{np} - A_0)^{-1}A_1 Y_{t-1} + (I_{np} - A_0)^{-1}\epsilon_t, \tag{10}$$

The difference between a standard VAR and QVAR model on the same shock is that QVAR measures the shock's effect on all the quantiles while standard VAR measures the shock to variable $i$ at $t$ on the conditional expectations only. In this study, QVAR is implemented to measure the dynamic connectedness among different countries by evaluating the average impact a shock in each country has on all other countries. QVAR reports a set of interconnectedness measures such as the total connectedness index (TCI) which measures overall interconnectedness among all countries; the "FROM" connectedness index represents the directional spillover received by the country $i$ from all other countries; the "TO" connectedness index represents the directional spillover sent by the country $i$ to all other countries; "Net Pairwise Directional Connectedness (NPDC)" measures the influence/spillover of country $i$ has on country $j$. If NPDC is positive, then the country $i$ dominates country $j$, otherwise, country $j$ dominates country $i$.

4. **Data**

To perform the cointegration, causality, and connectedness analyses, time series data are collected from various sources. Net inflows of Foreign Direct Investment (FDI), import of intermediate goods, technology readiness, innovation, and infrastructure data are obtained from The World Bank. The net inflows of FDI and intermediate imports are measured in US$. Technology is measured by ICT goods intermediate imports (% total goods intermediate imports) which is defined as intermediate imports of information and communication technology which include computers and peripheral equipment, communication equipment, consumer electronic equipment, electronic components, and other information and technology goods (miscellaneous). Innovation is measured by Patent applications from residents and non-residents. Infrastructure is measured by Industry (including construction), value added (constant 2015 US$) which is defined as industry (including construction) corresponds to ISIC divisions 05-43 and includes manufacturing (ISIC divisions 10-33). It comprises value added in mining, manufacturing (also reported as a separate subgroup), construction, electricity, water, and gas. The "value-added" is the net output of a sector after adding up all outputs and subtracting intermediate inputs. For Southeast Asian countries, we exclude Brunei, Burma, Cambodia, Laos, and Timor-Leste because of the lack of information on these countries. To examine the spillover effect, we also collect data in China. The descriptive results of the variables for 6 Southeast Asian countries are presented in Table TA1.

The cointegration, causality, and dynamic connectedness analyses are implemented by using the *vars*[8], *ConnectednessApproach*[9], *tseries*[10] packages from R. Table 1 shows that the variables for each country are non-stationary. For dynamic connectedness, we report the results using the total connected index (TCI) as a measure for the average impact a shock in each country has on all other countries, which can be broken into two directional connectedness. $TO_i$ represents the impacted county $i$ has on all other countries and $FROM_i$ reports the impact all other countries have on the country $i$. The differences between the total directional $TO_i$ and $FROM_i$ is given as $NET_i$, which is the net influence of the country $i$. The net pairwise directional connectedness ($NPDC_{i,j}$) measure is also reported. For a large network, $NPDC_{i,j}$ are computationally efficient as well as more accurate than *TCI* in reducing bias results.

Table 1. The results of ADF and KPSS tests on variables.

|  | FDI-inflow | Intermediate Imports | Innovation | Tech-Ready | Infrastructure |
|---|---|---|---|---|---|
| Indonesia | N* | N* | N* | N | N |
| Malaysia | N | N* | N* | N* | N |
| The Philippines | N* | N* | N* | N* | N |
| Singapore | N* | N* | N* | N | N |
| Thailand | N | N* | N | N | N* |
| Vietnam | N* | N* | N* | N* | N |

Stationarity (N: not supported; S: Supported); p-*value<=0.1, * p-value<=0.05, \*\* p-value<=0.01, \*\*\*, p-value<=0.005*

## 5. Empirical Findings and Discussion

In Table 2, Rolling regression reports the relationship among countries with the offshoring indicator intermediate imports as the dependent variable and other variables as independent variables.

Table 2a. Rolling regression results (mean $R^2$ values) for RQ1

| Country | Rolling Windows for RQ1 | | | | |
|---|---|---|---|---|---|
|  | 4 | 5 | 6 | 7 | 8 |
| Indonesia | 0.285363 | 0.384372 | 0.528852 | 0.530044 | 0.490365 |
| Malaysia | 0.156445 | 0.232372 | 0.325485 | 0.359102 | 0.38402 |

---

[8] https://cran.r-project.org/web/packages/vars/index.html
[9] https://cran.r-project.org/web/packages/ConnectednessApproach/index.html
[10] https://cran.r-project.org/web/packages/tseries/index.html

| Country | | | | | |
|---|---|---|---|---|---|
| The Philippines | 0.124631 | 0.07306 | 0.16612 | 0.18679 | 0.17125 |
| Singapore | 0.01274 | 0.0638 | 0.074861 | 0.107654 | 0.03904 |
| Thailand | 0.191474 | 0.46623 | 0.542668 | **0.592898** | **0.591953** |
| Vietnam | 0.560164 | 0.516059 | 0.461712 | 0.445153 | 0.46223 |

Table 2b. Rolling regression results (mean $R^2$ values) for RQ2

| Country | Rolling Windows for RQ2 | | | | |
|---|---|---|---|---|---|
| | 4 | 5 | 6 | 7 | 8 |
| Indonesia | 0.37929 | 0.574366 | **0.662362** | **0.635731** | **0.574498*** |
| Malaysia | 0.158862 | 4.63E-05 | 0.068598 | 0.087925 | 0.120546 |
| The Philippines | 0.07437 | 0.09214 | 0.008391 | 0.013977 | 0.03343 |
| Singapore | 0.069297 | 0.031426 | 0.052566 | 0.043459 | 0.003559 |
| Thailand | 0.04785 | 0.219438 | 0.193034 | 0.210782 | 0.14526 |
| Vietnam | 0.361676 | 0.544037 | **0.676735*** | **0.723162**** | **0.743472**** |

Table 2c. Rolling regression results (mean $R^2$ values) for RQ3

| Country | Rolling Windows for RQ3 | | | | |
|---|---|---|---|---|---|
| | 3 | 4 | 5 | 6 | 7 |
| Indonesia | 0.230579 | 0.326052 | 0.410945 | **0.514884** | **0.545939*** |
| Malaysia | 0.356324 | 0.332844 | 0.28204 | 0.323119 | 0.353667 |
| The Philippines | 0.234993 | 0.200471 | 0.254315 | 0.283863 | 0.311197 |
| Singapore | 0.064362 | 0.038049 | 0.000197 | 0.042536 | 0.106625 |
| Thailand | 0.134636 | 0.176175 | 0.249651 | 0.265452 | 0.251777 |
| Vietnam | 0.439613 | 0.555425 | 0.624676 | **0.640471** | **0.651207*** |

In the rolling regression analysis, the size of rolling windows must be larger than the number of variables in the model. Thus, we choose rolling windows from 4 years to 8 years for RQ1 and RQ2 and 3 years to 7 years for RQ3. Tables 2a to 2c report the $R^2$ values of rolling regression analysis to address RQ1, RQ2, and RQ3. Numbers in boldface have f-test *p*-values <=0.1, with * means *p*-values <=0.05, with ** means *p*-values <=0.01. Because the rolling regression model is sensitive to outliers and the prediction powers are affected by the rolling window size, the mean $R^2$ value in Tables 2a, 2b and 2c tend to be relatively low on the small rolling window size. The GFC in 2008 and the COVID-19 pandemic might affect the results of rolling regression as outliers. In addition, the performance of rolling regression heavily depends on the sample size. For RQ1, we observed that only Indonesia, Thailand, and Vietnam have high $R^2$ values when the size of the rolling window is large. For RQ2 and RQ3, only Indonesia and Vietnam have high $R^2$ values when the size of the rolling window is large. Interestingly, not all countries receive more FDI by investing in infrastructure/technology and countries might not experience improvements in innovation by investing in infrastructure/technology. Only Indonesia and Vietnam report a significant relationship between net FDI inflows and manufacturing outputs. Due to the limitation of the rolling regression model to answer RQ1, RQ2, and RQ3, a standard VAR model is tested on these variables and the results are reported in Tables 3 and 4. The results from VAR indicate that intermediate imports and net FDI inflows have a complementary relationship while net FDI inflows positively contribute to the value of intermediate imports of the receiving countries through increased capacity and productivity by the means of technology transfer, improved skills of the local workforce through training, and management capabilities. These findings in VAR indicate significant relationships among the factors for offshore decisions.

Table 3. Granger Causality results of offshore decision factors in Southeast Asia countries

| Country | | FDI | Intermediate imports | Infrastructure | Technology | Innovation |
|---|---|---|---|---|---|---|
| Indonesia | FDI | - | - | - | - | |
| | Intermediate imports | - | - | - | - | - |
| | Infrastructure | - | 4.3836** (0.03629) | | - | 7.5406** (0.01042) |
| | Technology | 4.3638** (0.02534) | 3.4921** (0.04818) | - | - | - |
| | Innovation | - | - | 5.2178** (0.03014) | - | - |
| Malaysia | FDI | - | - | - | 3.7163** (0.04069) | - |
| | Intermediate imports | - | - | 2.9842* (0.08408) | - | 2.578* (0.09866) |
| | Infrastructure | - | - | - | - | 9.2222*** (0.00089) |
| | Technology | 7.3499*** (0.00359) | 8.06*** (0.00236) | 8.72*** (0.00116) | - | 5.2265** (0.01048) |
| | Innovation | - | - | 3.4599** (0.04142) | 4.0859** (0.02482) | - |
| The Philippines | FDI | - | 12.238*** (0.00158) | 3.8596** (0.03799) | - | 10.51** (0.04878) |
|  | Intermediate imports | - | - | 6.619** (0.01568) | - | 5.0669** (0.01549) |
|  | Infrastructure | - | - | - | 4.8933** (0.01336) | 4.8712** (0.01773) |
|  | Technology | - | - | - | - | - |
|  | Innovation | - | - | - | - | - |
| Singapore | FDI | - | 4.9673** (0.01659) | 4.1719** (0.0291) | 3.6783** (0.03453) | 3.1312* (0.0877) |
|  | Intermediate imports | - | - | - | - | - |
|  | Infrastructure | 2.7073* (0.08889) | - | - | - | - |
|  | Technology | 5.5842*** (0.00813) | 3.1284* (0.05501) | 7.7004*** (0.00209) | - | - |
|  | Innovation | - | - | - | 3.388** (0.04402) | - |
| Thailand | FDI | - | - | - | - | - |
|  | Intermediate imports | - | - | - | - | 3.879** (0.05886) |
|  | Infrastructure | - | - | - | - | 4.7974** (0.01435) |
|  | Technology | - | - | 3.3492** (0.05373) | - | 6.4962*** (0.00440) |
|  | Innovation | - | - | - | 2.5379* (0.09323) | - |
| Vietnam | FDI | - | - | - | - | - |
|  | Intermediate imports | 3.2974* (0.08011) | - | 6.1509*** (0.00553) | 4.6928** (0.03895) | - |
|  | Infrastructure | 3.2116* (0.08393) | - | - | - | - |
|  | Technology | - | 5.2384** (0.02984) | - | - | - |
|  | Innovation | 4.9654** (0.01661) | - | - | - | 5.543** (0.0258) |

Johansen test is used to assess the validity of cointegration relationships in Table 4. Since we only have two univariate time series, we can only have two ranks: r=0 or r<=1. This means that either there is no cointegration (r=0) or there is (r<=1).

The hypothesis is stated as:

$H_0$: no cointegration

$H_1$: $H_0$ does not hold

The first hypothesis, r = 0, tests for the presence of cointegration. Thus, a footnote as Trace statistics of r = 0 is reported in Table 4 (*** at 1%, ** at 5%, and * at 10% levels of significance). For instance, in the FDI and Technology example for Indonesia, since the test statistic exceeds the 10% level (18.85>17.85) we have strong evidence to reject the null hypothesis of no cointegration. Therefore, it is reported in Table to the trace statistic 18.85 with *.

Table 4. Cointegration results of offshore decision factors in Southeast Asia countries

| Country | | FDI | Intermediate imports | infrastructure | Technology | Innovation |
|---|---|---|---|---|---|---|
| Indonesia | FDI | | | | | |
| | Intermediate imports | - | | | | |
| | Infrastructure | - | - | | | |
| | Technology | 18.85* | - | - | | |
| | Innovation | - | 18.75* | - | 18.24* | |
| | | | | | | |
| Malaysia | FDI | | | | | |
| | Intermediate imports | - | | | | |
| | Infrastructure | 21.37** | 20.77** | | | |
| | Technology | - | - | 27.06*** | | |
| | Innovation | - | 26.47** | 31.63*** | 41.10*** | |
| | | | | | | |
| The Philippines | FDI | | | | | |
| | Intermediate imports | - | | | | |
| | Infrastructure | 18.50* | 18.38* | | | |
| | Technology | - | 18.68* | 34.85*** | | |
| | Innovation | 25.02*** | 22.79** | 30.92*** | 25.86*** | |
| | | | | | | |
| Singapore | FDI | | | | | |
| | Intermediate imports | - | | | | |
| | Infrastructure | - | - | | | |
| | Technology | 33.15*** | 24.97*** | 37.39*** | | |
| | Innovation | - | - | - | 21.50** | |
| | | | | | | |
| Thailand | FDI | | | | | |
| | Intermediate imports | - | | | | |
| | Infrastructure | 18.34* | - | | | |
| | Technology | 19.42* | - | - | | |
| | Innovation | - | - | 23.00** | 25.24** | |
| | | | | | | |
| Vietnam | FDI | | | | | |

|  | Intermediate imports | 28.48*** | | | |
|---|---|---|---|---|---|
|  | Infrastructure | - | 21.85** | | |
|  | Technology | 20.87** | 18.47* | - | |
|  | Innovation | 18.64* | 29.60*** | 21.26** | - |

Tables 3 and 4 summarize the causality and co-integration relations between technology readiness, infrastructure, innovation, net FDI inflows, and intermediate import variables. The relationship between innovation and technology readiness has drawn research interests since the second half of the 20[th] century as there are other factors determining the nature of this relationship such as demand and competition conditions [44]. In Southeast countries, there is evidence to support co-integration between technology and innovation.

To assess the spillover of offshore decision factors and the dynamic interconnected relationship between China and 6 Southeast Asia countries (RQ4 & RQ5), Tables 5-10 reports the findings of static spillover connectedness for three quantiles such as extreme lower 0.1, lower 0.25, and intermediate 0.5 on offshoring decisions factors such as intermediate imports and net FDI inflows.

Table 5. Dynamic connectedness analysis on offshoring—intermediate imports with quantile=0.1

|  | China | Indonesia | Malaysia | The Philippines | Singapore | Thailand | Vietnam | Contribution "FROM" others |
|---|---|---|---|---|---|---|---|---|
| China | 20.09 | 17.49 | 11.69 | 4.28 | 15.62 | 20.29 | 10.53 | 79.91 |
| Indonesia | 25.27 | 12.52 | 8.29 | 18.02 | 9.39 | 20.47 | 6.04 | 87.48 |
| Malaysia | 25.61 | 14.65 | 6.48 | 13.31 | 10.99 | 21.9 | 7.06 | **93.52** |
| The Philippines | 26.73 | 15.79 | 4.25 | 11.41 | 11.28 | 23.74 | 6.8 | 88.59 |
| Singapore | 25.28 | 14.23 | 7.13 | 12.45 | 11.43 | 22.02 | 7.45 | 88.57 |
| Thailand | 24.27 | 16.45 | 6.87 | 7.25 | 13.64 | 22.79 | 8.73 | 77.21 |
| Vietnam | 25.08 | 12.66 | 8.48 | 14.94 | 10.55 | 20.94 | 7.35 | 92.65 |
| Contribution "TO" Others | **152.24** | 91.27 | 46.7 | 70.25 | 71.48 | 129.37 | 46.6 | 607.94 |
| TCI-Total Contribution Including Own | **172.33** | 103.8 | 53.18 | 81.66 | 82.91 | 152.17 | 53.96 |  |
| NET-Net Spillovers | **72.33** | 3.8 | -46.82 | -18.34 | -17.09 | 52.17 | -46.04 |  |
| Total NPDC | **6** | 3 | 1 | 4 | 2 | 5 | 0 |  |

Table 6. Dynamic connectedness analysis on offshoring—intermediate imports with quantile=0.25

|  | China | Indonesia | Malaysia | The Philippines | Singapore | Thailand | Vietnam | Contribution "FROM" others |
|---|---|---|---|---|---|---|---|---|
| China | 22.58 | 9.14 | 10.62 | 27.66 | 7.9 | 16.71 | 5.39 | 77.42 |
| Indonesia | 16.17 | 18.5 | 16.44 | 4.12 | 16.26 | 18.11 | 10.4 | 81.5 |
| Malaysia | 14.15 | 17.16 | 19.6 | 6.67 | 15.96 | 16.38 | 10.08 | 80.4 |
| The Philippines | 13.12 | 15.58 | 21.7 | 5.12 | 17.13 | 15.99 | 11.37 | **94.88** |
| Singapore | 12.66 | 16.4 | 21.95 | 8.16 | 15.76 | 14.86 | 10.19 | 84.24 |
| Thailand | 12.1 | 12.6 | 24.31 | 17.49 | 12.89 | 12.15 | 8.46 | 87.85 |
| Vietnam | 15.91 | 18.26 | 16.84 | 4.47 | 16.19 | 18.37 | 9.96 | 90.04 |
| Contribution "TO" Others | 84.12 | 89.13 | **111.85** | 68.57 | 86.33 | 100.42 | 55.89 | 596.32 |
| TCI-Total Contribution Including Own | 106.7 | 107.63 | **131.46** | 73.69 | 102.09 | 112.57 | 65.86 |  |
| NET-Net Spillovers | 6.7 | 7.63 | **31.46** | -26.31 | 2.09 | 12.57 | -34.14 |  |
| Total NPDC | 4 | 4 | 4 | 2 | 2 | 4 | 1 |  |

Table 7. Dynamic connectedness analysis on offshoring—intermediate imports with quantile=0.5

|  | China | Indonesia | Malaysia | The Philippines | Singapore | Thailand | Vietnam | Contribution "FROM" others |
|---|---|---|---|---|---|---|---|---|
| China | 8.17 | 1.61 | 54.14 | 15.19 | 14.15 | 5.07 | 1.68 | 91.83 |
| Indonesia | 9.74 | 3.81 | 44.59 | 10.78 | 16 | 9.99 | 5.09 | 96.19 |
| Malaysia | 7.3 | 1.53 | 53.46 | 14.28 | 15.12 | 6.31 | 2.01 | 46.54 |
| The Philippines | 34.88 | 19 | 14.12 | 9.11 | 5.78 | 8.35 | 8.74 | 90.89 |
| Singapore | 3.54 | 0.71 | 53.09 | 14.65 | 17.74 | 6.72 | 3.57 | 82.26 |
| Thailand | 7.77 | 0.9 | 52.63 | 13.61 | 15.68 | 5.94 | 3.47 | 94.06 |
| Vietnam | 7.43 | 1.34 | 52.18 | 12.24 | 16.58 | 6.82 | 3.41 | **96.59** |
| Contribution "TO" Others | 70.66 | 25.08 | **270.75** | 80.75 | 83.32 | 43.27 | 24.56 | 598.37 |
| TCI-Total Contribution Including Own | 78.82 | 28.89 | **324.2** | 89.86 | 101.05 | 49.2 | 27.97 |  |
| NET-Net Spillovers | -21.18 | -71.11 | **224.2** | -10.14 | 1.05 | -50.8 | -72.03 |  |
| Total NPDC | 4 | 1 | **5** | 4 | 4 | 2 | 1 |  |

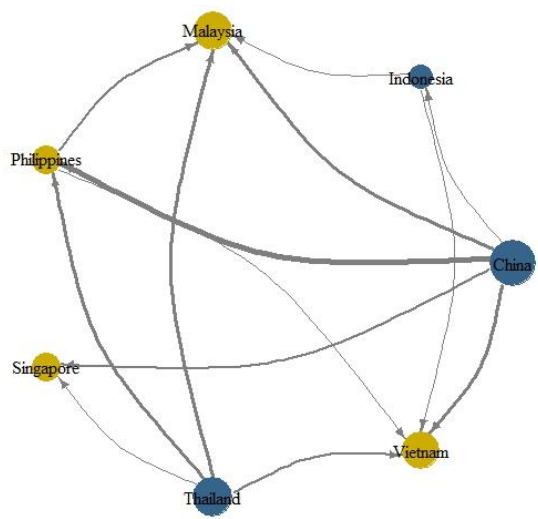

Figure 3. NPDC measure plot quantile=0.10 on immediate imports

Tables 5-7 showed some interesting results of static spillover connectedness in terms of intermediate imports: a) When the quantile is extremely low (0.1), the spillover contributions from (to) other countries vary between 93.51% (152.24%) to 77.21% (46.6%); b) When the quantile is relatively low (0.25), the spillover contributions from (to) other countries vary between 94.88% (111.88%) to 77.42% (55.89%); c) For the intermediate quantile (0.5), Vietnam has the highest contribution (96.89%) from other countries but the lowest contribution (24.56%) to them, while Malaysia has the lower contribution (46.54%) from other countries but an extremely high contribution (270.75%) to others; d) The from/to bidirectional spillover suggested that fluctuations of intermediate imports in one country will highly impact other countries in the region. The connectedness in the last two decades among these countries has intensified. The net spillover patterns of intermediate imports indicated that all countries are contributors and recipients of offshore decisions, and the findings show which country receives more spillover or transfers more from/to others. At low and intermediate quantiles, Malaysia transfers more spillover to other countries and Vietnam receives more spillover from other countries.

However, we are more interested in the dynamics of the net pairwise directional connectedness (NPDC) of spillover throughout the counties in the whole system. Figure 3 shows the strengths and directions of a net spillover over time with an extreme low quantile (0.1). We also report the results of other quantiles (0.25 and 0.5 in Appendix Figures A1 and A2). In Figure 3, the arrow pointing from one country to another defines positive net spillovers. The thickness of the arrow defines the strength of spillover. The color of the country node defines the nature of contributors and recipients. The NPDC figures show that the network of net pairwise directional spillovers from intermediate import among countries differs across quantiles. To evaluate the sensitivity of dynamic connectedness analysis on intermediate import among these countries, we examine the net directional accuracy of each country in the system across a sequel of quantiles (0.05, 0.1, 0.15, 0.20. …0.90, 0.95). The results are reported in Appendix Tables A2 and A3. The total spillover contribution of intermediate imports is very close across the sequel of quantiles, but the standard deviation (STDEV) of spillover contribution to other countries in Table A3 is greater than from other countries in Table A2. These findings show the dynamic interconnection within the system is strong on intermediate imports. When the quantile is 0.5, there is the highest standard deviation among the countries. When the quantile value is lower than 0.45 or higher than 0.70, there is a stable pattern of spillover contribution among countries.

Table 8. Dynamic connectedness analysis (quantile=0.1) on offshoring—net FDI inflows

|  | China | Indonesia | Malaysia | The Philippines | Singapore | Thailand | Vietnam | Contribution "FROM" others |
|---|---|---|---|---|---|---|---|---|
| China | 22.26 | 15.29 | 3.65 | 8.6 | 5.01 | 21.3 | 23.9 | 77.74 |
| Indonesia | 11.97 | 14.7 | 9.94 | 29.7 | 6.41 | 14.74 | 12.54 | 85.3 |
| Malaysia | 18.21 | 6.2 | 8.6 | 22.9 | 8.88 | 17.68 | 17.53 | 91.4 |
| The Philippines | 2.1 | 3.38 | 18.64 | 63.77 | 4.94 | 1.97 | 5.2 | 36.23 |
| Singapore | 0.54 | 7.29 | 20.44 | 60.32 | 9 | 1.73 | 0.69 | 91 |
| Thailand | 15.75 | 7.34 | 9.67 | 34.37 | 6.96 | 13.93 | 11.99 | 86.07 |
| Vietnam | 4.7 | 1.83 | 16.36 | 54.82 | 8.44 | 8.19 | 5.66 | **94.34** |
| Contribution "TO" Others | 53.27 | 41.32 | 78.7 | **210.7** | 40.64 | 65.6 | 71.84 | 562.09 |
| TCI-Total Contribution Including Own | 75.53 | 56.02 | 87.3 | **274.47** | 49.64 | 79.53 | 77.5 |  |
| NET-Net Spillovers | -24.47 | -43.98 | -12.7 | **174.47** | -50.36 | -20.47 | -22.5 |  |
| Total NPDC | 1 | 2 | 2 | **6** | 3 | 3 | 4 |  |

Table 9. Dynamic connectedness analysis (quantile=0.25) on offshoring—net FDI inflows

|  | China | Indonesia | Malaysia | The Philippines | Singapore | Thailand | Vietnam | Contribution "FROM" others |
|---|---|---|---|---|---|---|---|---|
| China | 8.6 | 8.22 | 13.79 | 48.5 | 4.29 | 9.56 | 7.05 | 91.4 |
| Indonesia | 6.63 | 13.82 | 13.73 | 48.97 | 3.53 | 9.17 | 4.16 | 86.18 |
| Malaysia | 16.57 | 14.77 | 8.41 | 27.7 | 5.96 | 15 | 11.59 | 91.59 |
| The Philippines | 12.7 | 2.79 | 12.26 | 52.56 | 1.62 | 12.57 | 5.5 | 47.44 |
| Singapore | 4.55 | 6.2 | 16.01 | 61.69 | 3.29 | 6.37 | 1.89 | 96.71 |
| Thailand | 18.18 | 7.81 | 11.5 | 41.54 | 2.12 | 8.85 | 10 | 91.15 |
| Vietnam | 6.06 | 1.15 | 16.61 | 64.88 | 2.82 | 7.21 | 1.27 | **98.73** |
| Contribution "TO" Others | 64.69 | 40.94 | 83.9 | **293.27** | 20.33 | 59.88 | 40.18 | 603.19 |
| TCI-Total Contribution Including Own | 73.28 | 54.76 | 92.31 | **345.83** | 23.63 | 68.73 | 41.45 | |
| NET-Net Spillovers | -26.72 | -45.24 | -7.69 | **245.83** | -76.37 | -31.27 | -58.55 | |
| Total NPDC | 3 | 3 | 2 | **6** | 1 | 3 | 3 | |

Table 10. Dynamic connectedness analysis (quantile=0.5) on offshoring—net FDI inflows

| | China | Indonesia | Malaysia | The Philippines | Singapore | Thailand | Vietnam | Contribution "FROM" others |
|---|---|---|---|---|---|---|---|---|
| China | 33.8 | 8.35 | 8.86 | 29.03 | 4.54 | 7.81 | 7.62 | 66.2 |
| Indonesia | 22.67 | 31.44 | 6.96 | 16.92 | 2.63 | 15.54 | 3.85 | 68.56 |
| Malaysia | 31.26 | 21.79 | 4.64 | 22.07 | 3.12 | 8.31 | 8.8 | 95.36 |
| The Philippines | 22.2 | 26.77 | 6.65 | 23.07 | 2.94 | 13.16 | 5.21 | 76.93 |
| Singapore | 18.62 | 39.2 | 8.19 | 9.7 | 2.81 | 18.97 | 2.5 | **97.19** |
| Thailand | 13.28 | 48.3 | 5.53 | 7.92 | 3.8 | 17.6 | 3.57 | 82.4 |
| Vietnam | 26.06 | 22.51 | 5.79 | 26.5 | 2.06 | 11.19 | 5.88 | 94.12 |
| Contribution "TO" Others | 134.09 | **166.93** | 41.98 | 112.15 | 19.09 | 75 | 31.54 | 580.77 |
| TCI-Total Contribution Including Own | 167.89 | **198.36** | 46.62 | 135.21 | 21.91 | 92.59 | 37.42 | |
| NET-Net Spillovers | 67.89 | **98.36** | -53.38 | 35.21 | -78.09 | -7.41 | -62.58 | |
| Total NPDC | 5 | **5** | 1 | 4 | 0 | 4 | 2 | |

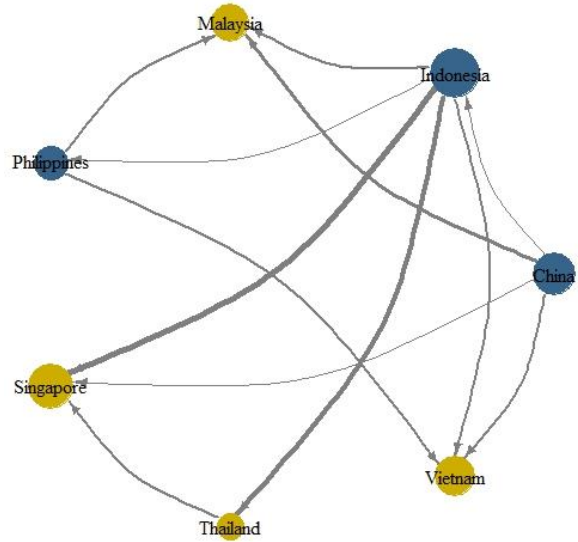

Figure 4. NPDC measure plot quantile=0.5 based on net FDI inflows

Tables 8-10 show interesting results of static spillover connectedness in terms of net FDI inflows: a) The Philippines has the highest value of spillover to other countries at quantile=0.1 and the lowest value of spillover received from other countries; b) When the quantile is extreme lower (0.1), the contributions from (to) others vary between 94.34% (210.7%) to 36.23% (40.64%); c) With lower quantile (0.25), the contributions from (to) others vary between 98.73% (293.27%) to 47.44% (20.33%); d) For the intermediate quantile (0.5), Singapore has the highest contribution (97.19%) from others but the lowest contribution (19.09%) to others while China has a lower contribution (66.2%) from others and Indonesia has the highest contribution (166.93%) to others; e) The from/to bidirectional spillover suggested that effects of Net FDI inflows fluctuations in one country will highly impact other countries in the region.

The connectedness in the last two decades among these countries on the net FDI inflows has intensified. The net spillover patterns of intermediate import indicated that all countries are contributors and recipients of net FDI inflows and the findings show which country receives more spillover or transfers more from/to others. At the extremely low quantile, The Philippines transfers more spillover to other countries and Vietnam receives more spillover from other countries. Figure 4 shows the dynamics of the net pairwise directional connectedness (NPDC) of spillover throughout the counties in the whole system. In Figure 4, the strengths and directions of net spillover over time with extremely lower quantiles (0.5) are reported. We also report the results of other quantiles (0.1 and 0.25 in appendix Figures A3 and A4)  These NPDC figures show that the network of net pairwise directional spillovers for the net FDI inflows among countries differs across the quantiles as becomes sparse at extreme low and low quantiles. The spillover effect on net

FDI inflows might be caused by the countries outside the system. To evaluate the sensitivity of dynamic connectedness analysis on intermediate import among these countries, we examine the net directional accuracy of each country in the system across a sequel of quantiles (0.05, 0.1, 0.15, 0.20. …0.90, 0.95). These results are reported in Appendix Tables A4 and A5. The total spillover contribution of the net FDI inflows from (to) other countries is very close across the sequel of quantiles, but the standard deviation (STDEV) of spillover contribution to other countries (In Table A5) is greater than from other countries (in Table A4). These findings confirm a strong dynamic interconnection within the system in terms of net FDI inflows. However, the standard deviation for the net FDI inflows shows at quantile=0.35 their highest standard deviation. When the quantile value is lower than 0.30 or higher than 0.70, there is a stable pattern of spillover contribution among countries (Figure 5).

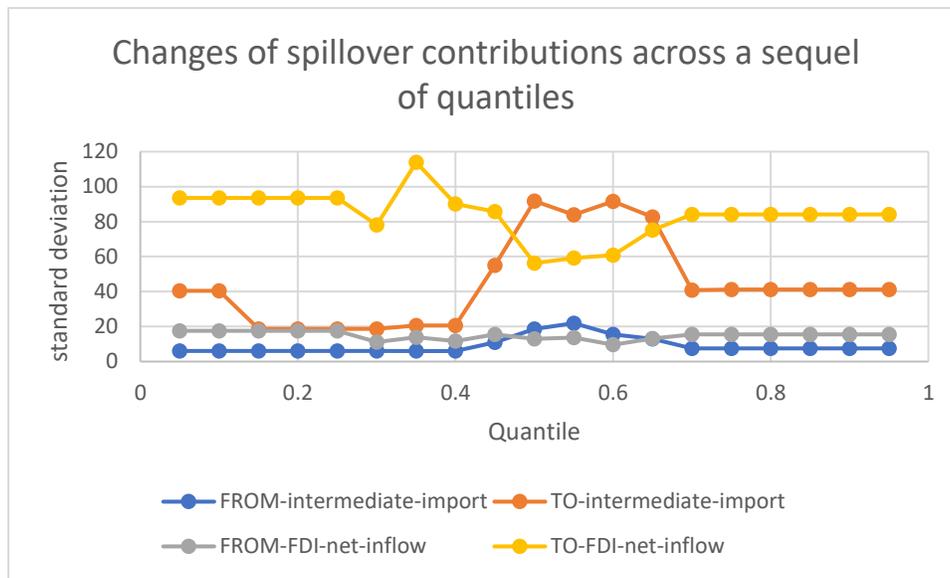

Figure 5. Changes of spillover contributions across a sequel of quantiles

### 6. Managerial implications

The cointegration, causality, and dynamic connectedness analyses show a multi-way cointegration and causal relationship among infrastructure, innovation, technology readiness, intermediate imports and intermediate imports, and dynamic connectedness relationship among Southeast Asian countries between 2000-2020. The cointegration and causality analyses suggest a strong multi-way cointegration and causal relationship among import, export, and FDI variables. They also indicate that countries that invested in infrastructure/technology saw more innovation, and in turn, received more FDI. Countries that receive more FDI present strong cointegration and causal relationships with intermediate imports and intermediate imports, which are the outcomes of the offshoring decision. FDI helps these countries to recover from the headwind of these events after GCP.

Among the seven countries studied, China is the largest trading partner of Indonesia, Malaysia, Singapore, and Vietnam. It is also the second largest trading partner The Philippines and Thailand. The relocation of manufacturing and diversification of the supply chain out of China have spillover effects to the Southeast Asia countries based on the results of cointegration and causality analysis results.

The dynamic connectedness analysis shows the path of spillover among the countries and reveals the positive interconnected relationship between China and Southeast Asia countries. For both offshore decision factors of intermediate import and net FDI inflows, the overall spillover at different quantile values shows the pattern of bidirectional spillover between China and other Southeast Asia countries. However, the asymmetric spillover effects show some countries contribute more, and such differences among countries are also found in the standard cointegration and causality analysis.

## 7. Conclusions

This study examines bidirectional relationships for three pairs of cointegration between FDI and innovation, FDI and competitiveness, innovation, and competitiveness. The empirical results on causality and cointegration provide several insights for policymakers. First, policy decisions solely based on the correlation of two variables can be misleading due to statistical noise consisting of residuals and errors in measurement and sampling. Second, co-integration and Granger-causes indicate that FDI enhances intermediate imports and, in return increases innovations, infrastructure, and technology-readiness. Third, our findings show that it is important to pay close attention to the network structure and dynamic interconnection of spillovers among the studied countries. Policymakers might use our framework to help businesses on offshore decisions and find the best candidate countries to relocate their factories. There are limitations in our study mainly due to the availability of data. First, other factors such as the market and trade openness index, capital account openness index, and other socioeconomic indices can be added to the model in future studies when the data are available in the studied countries. Second, the spillover effect from other developed countries such as Japan and the USA might be important in terms of FDI.

**References**

1. B. Schmeisser. A Systematic Review of Literature on Offshoring of Value Chain Activities, Journal of International Management, Volume 19, Issue 4, 2013, Pages 390-406, ISSN 1075-4253, https://doi.org/10.1016/j.intman.2013.03.011.
2. Fifarek BJ, Veloso FM, Davidson CI. Offshoring technology innovation: A case study of rare-earth technology. Journal of Operations Management. 2008 2008/03/01/;26(2):222-238. doi: https://doi.org/10.1016/j.jom.2007.02.013.
3. Liu X, Yeung ACL, Lo CKY, et al. The moderating effects of knowledge characteristics of firms on the financial value of innovative technology products. Journal of Operations Management. 2014 2014/03/01/;32(3):79-87. doi: https://doi.org/10.1016/j.jom.2013.11.003.

4. Tate WL, Ellram LM, Schoenherr T, et al. Global competitive conditions driving the manufacturing location decision. Business Horizons. 2014 2014/05/01/;57(3):381-390. doi: https://doi.org/10.1016/j.bushor.2013.12.010.
5. Hult GTM. A focus on international competitiveness. Journal of the Academy of Marketing Science. 2012 2012/03/01;40(2):195-201. doi: 10.1007/s11747-011-0297-7.
6. Jensen PDØ, Pedersen T. Offshoring and international competitiveness: antecedents of offshoring advanced tasks. Journal of the Academy of Marketing Science. 2012 2012/03/01;40(2):313-328. doi: 10.1007/s11747-011-0286-x.
7. Mudambi R, Venzin M. The Strategic Nexus of Offshoring and Outsourcing Decisions. Journal of Management Studies. 2010 2010/12/01;47(8):1510-1533. doi: https://doi.org/10.1111/j.1467-6486.2010.00947.x.
8. Lewin AY, Massini S, Peeters C. Why are companies offshoring innovation? The emerging global race for talent. Journal of International Business Studies. 2009 2009/08/01;40(6):901-925. doi: 10.1057/jibs.2008.92.
9. Massini S, Perm-Ajchariyawong N, Lewin AY. Role of Corporate-Wide Offshoring Strategy on Offshoring Drivers, Risks and Performance. Industry and Innovation. 2010 2010/08/01;17(4):337-371. doi: 10.1080/13662716.2010.496242.
10. Markusen JR. Modeling the offshoring of white-collar services: from comparative advantage to the new theories of trade and FDI. National Bureau of Economic Research Cambridge, Mass., USA; 2005.
11. Gugler P, Brunner S. FDI Effects on National Competitiveness: A Cluster Approach. International Advances in Economic Research. 2007 2007/08/01;13(3):268-284. doi: 10.1007/s11294-007-9091-1.
12. Dunning JH, Zhang FJTc. Foreign direct investment and the locational competitiveness of countries. 2008;17(3):1.
13. Prime PB, Subrahmanyam V, Lin CM. Competitiveness in India and China: the FDI puzzle. Asia Pacific Business Review. 2012 2012/07/01;18(3):303-333. doi: 10.1080/13602381.2011.605673.
14. Amendolagine V, Capolupo R, Ferri G. Innovativeness, offshoring and black economy decisions. Evidence from Italian manufacturing firms. International Business Review. 2014 2014/12/01/;23(6):1153-1166. doi: https://doi.org/10.1016/j.ibusrev.2014.03.011.
15. Massini S, Miozzo M. Outsourcing and Offshoring of Business Services: Challenges to Theory, Management and Geography of Innovation. Regional Studies. 2012 2012/10/01;46(9):1219-1242.
16. Baum, Christopher, Lööf, Hans, Stephan, Andreas and Viklund-Ros, Ingrid, (2021), The impact of offshoring on productivity and innovation: Evidence from Swedish manufacturing firms, No 1014, Boston College Working Papers in Economics, Boston College Department of Economics
17. Amendolagine V, Capolupo R, Ferri G. Innovativeness, offshoring and black economy decisions. Evidence from Italian manufacturing firms. International Business Review. 2014
18. Thomson R. National scientific capacity and R&D offshoring. Research Policy. 2013 2013/03/01/;42(2):517-528. doi: https://doi.org/10.1016/j.respol.2012.07.003.
19. Gersbach H, Schmutzler A. Foreign direct investment and R&D-offshoring. Oxford Economic Papers. 2011;63(1):134-157. doi: 10.1093/oep/gpq024.
20. Chung, S. (2011). Innovation, competitiveness, and growth: Korean experiences. *ABCDE*, *17*, 333.
21. OECD iLibrary https://www.oecd.org/daf/inv/investment-policy/2487495.pdf
22. Chrid N, Saafi S, Chakroun M. Export Upgrading and Economic Growth: A Panel Cointegration and Causality Analysis. Journal of the Knowledge Economy. 2021 2021/06/01;12(2):811-841. doi: 10.1007/s13132-020-00640-6.
23. da Silveira GJC. An empirical analysis of manufacturing competitive factors and offshoring. International Journal of Production Economics. 2014 2014/04/01/; 150:163-173. doi: https://doi.org/10.1016/j.ijpe.2013.12.031.
24. Bock S. Supporting offshoring and nearshoring decisions for mass customization manufacturing processes. European Journal of Operational Research. 2008 2008/01/16/;184(2):490-508. doi: https://doi.org/10.1016/j.ejor.2006.11.019.
25. Gray JV, Esenduran G, Rungtusanatham MJ, et al. Why in the world did they reshore? Examining small to medium-sized manufacturer decisions. Journal of Operations Management. 2017 2017/03/01/;49-51:37-51. doi: https://doi.org/10.1016/j.jom.2017.01.001.
26. Ishizaka A, Bhattacharya A, Gunasekaran A, et al. Outsourcing and offshoring decision making. International Journal of Production Research. 2019 2019/07/03;57(13):4187-4193. doi: 10.1080/00207543.2019.1603698.

27. Jahns C, Hartmann E, Bals L. Offshoring: Dimensions and diffusion of a new business concept. Journal of Purchasing and Supply Management. 2006 2006/07/01/;12(4):218-231. doi: https://doi.org/10.1016/j.pursup.2006.10.001.
28. Johansson M, Olhager J. Comparing offshoring and backshoring: The role of manufacturing site location factors and their impact on post-relocation performance. International Journal of Production Economics. 2018 2018/11/01/;205:37-46. doi: https://doi.org/10.1016/j.ijpe.2018.08.027.
29. Johansson M, Olhager J, Heikkilä J, et al. Offshoring versus backshoring: Empirically derived bundles of relocation drivers, and their relationship with benefits. Journal of Purchasing and Supply Management. 2019 2019/06/01/;25(3):100509. doi: https://doi.org/10.1016/j.pursup.2018.07.003.
30. Mihalache M, Mihalache OR. A Decisional Framework of Offshoring: Integrating Insights from 25 Years of Research to Provide Direction for Future* [https://doi.org/10.1111/deci.12206]. Decision Sciences. 2016 2016/12/01;47(6):1103-1149. doi: https://doi.org/10.1111/deci.12206.
31. Musteen M. Behavioral factors in offshoring decisions: A qualitative analysis. Journal of Business Research. 2016 2016/09/01/;69(9):3439-3446. doi: https://doi.org/10.1016/j.jbusres.2016.01.042.
32. Rahman HU, Raza M, Afsar P, et al. Empirical Investigation of Influencing Factors Regarding Offshore Outsourcing Decision of Application Maintenance. IEEE Access. 2021;9:58589-58608. doi: 10.1109/ACCESS.2021.3073315.
33. Tate WL. Offshoring and reshoring: U.S. insights and research challenges. Journal of Purchasing and Supply Management. 2014 2014/03/01/;20(1):66-68. doi: https://doi.org/10.1016/j.pursup.2014.01.007.
34. Goh SK, Wong KN. Malaysia's outward FDI: The effects of market size and government policy. Journal of Policy Modeling. 2011 2011/05/01/;33(3):497-510. doi: https://doi.org/10.1016/j.jpolmod.2010.12.008.
35. Bilgili F, Tülüce NS, Doğan I, et al. The causality between FDI and sector-specific production in Turkey: evidence from threshold cointegration with regime shifts. Applied Economics. 2016 2016/01/26;48(5):345-360. doi: 10.1080/00036846.2015.1078451.
36. Boateng A, Hua X, Nisar S, et al. Examining the determinants of inward FDI: Evidence from Norway. Economic Modelling. 2015 2015/06/01/;47:118-127. doi: https://doi.org/10.1016/j.econmod.2015.02.018.
37. Kinkel S, Maloca S. Drivers and antecedents of manufacturing offshoring and backshoring—A German perspective. Journal of Purchasing and Supply Management. 2009 2009/09/01/;15(3):154-165. doi: https://doi.org/10.1016/j.pursup.2009.05.007
38. Udemba EN, Yalçıntaş S. Interacting force of foreign direct invest (FDI), natural resource and economic growth in determining environmental performance: A nonlinear autoregressive distributed lag (NARDL) approach. Resources Policy. 2021 2021/10/01/;73:102168. doi: https://doi.org/10.1016/j.resourpol.2021.102168.
39. Herzer D, Hühne P, Nunnenkamp P. FDI and Income Inequality—Evidence from Latin American Economies [https://doi.org/10.1111/rode.12118]. Review of Development Economics. 2014 2014/11/01;18(4):778-793. doi: https://doi.org/10.1111/rode.12118.
40. Basu, P. and Chakraborty, C. and Reagle, D. (2003) 'Liberalization, FDI and growth in developing countries: a panel cointegration approach.', *Economic Inquiry.*, 41 (3). pp. 510-516.
41. Chakraborty C. and Nunnenkamp P. (2008) Economic Reforms, FDI, and Economic Growth in India: A Sector Level Analysis, World Development, 36 (7), pp. 1192-1212, https://doi.org/10.1016/j.worlddev.2007.06.014.
42. Plata, T. and Castaño, C. Analysis of the offshoring network: Empirical evidence of the implied comparative advantage in offshoring. The Journal of International Trade & Economic, 2021.
43. Canals, C. and Şener, F. (2014), Offshoring and intellectual property rights reform, Journal of Development Economics, 108, issue C, p. 17-31.
44. Feenstra, R. and Hanson, G. (1999), The Impact of Outsourcing and High-Technology Capital on Wages: Estimates For the United States, 1979–1990, The Quarterly Journal of Economics, 114, issue 3, p. 907-940
45. Dutta, A., & Roy, R. (2005). Offshore Outsourcing: A Dynamic Causal Model of Counteracting Forces. Journal of Management Information Systems, 22(2), 15–35. http://www.jstor.org/stable/40398743
46. Usman K, Bashir U. The causal nexus between intermediate imports and economic growth in China, India and G7 countries: granger causality analysis in the frequency domain. Heliyon. 2022 Aug 10;8(8):e10180. doi: 10.1016/j.heliyon.2022.e10180. PMID: 36016525; PMCID: PMC9396639.
47. Feng, Y., Wang, G-J., Zhu, Y., and Xie, C. (2023), "Systemic risk spillovers and the determinants in the stock markets of the Belt and Road countries", Emerging Markets Review, vol.55, doi: 10.17632/c9t5swh4vz.1

48. Diebold, F. X., & Yilmaz, K. (2009). Measuring financial asset return and volatility spillovers, with application to global equity markets. Economic Journal, 119, 158–171
49. Diebold, F.X., Yilmaz, K., 2012. Better to give than to receive: predictive directional measurement of volatility spillovers. Int. J. Forecast. 28 (1), 57–66.
50. Øystein Moen, Tord Tvedten & Andreas Wold | (2018) Exploring the relationship between competition and innovation in Norwegian SMEs, Cogent Business & Management, 5:1, 1564167, DOI: 10.1080/23311975.2018.1564167
51. Tiwari, A. K., and Abakah, E. J. A., Adewuyi, A. O., and Lee, C. C. Quantile Risk Spillovers and Interrelatedness between Energy and Agricultural Commodity Markets Using Realized Variance. Available at http://dx.doi.org/10.2139/ssrn.4014291
52. Lin B, Su T. Does COVID-19 open a Pandora's box of changing the connectedness in energy commodities? Res Int Bus Finance. 2021 Apr;56:101360. doi: 10.1016/j.ribaf.2020.101360. Epub 2020 Nov 27. PMID: 36540766; PMCID: PMC9756042.
53. Antonakakis, N., Gabauer, D., & Gupta, R. (2019). Greek economic policy uncertainty: Does it matter for Europe? Evidence from a dynamic connectedness decomposition approach. Physica A-statistical Mechanics and Its Applications, 535, 122280.
54. Gabauer, D., Chatziantoniou, I., and Stenfors, A. Model-free connectedness measures, Finance Research Letters, vol. 54, 2023, 103804, ISSN 1544-6123, https://doi.org/10.1016/j.frl.2023.103804.
55. Johansen, S.: 'Estimation and Hypothesis Testing of Cointegration Vectors in Gaussian Vector Autoregressive Models', Econometrica, 1991, 59, (6), pp. 1551-1580
56. Schüler, Y.S.: 'Asymmetric effects of uncertainty over the business cycle: A quantile structural vector autoregressive approach', 2014

**Appendix**

Table TA1 Descriptive results of annual data for 6 Southeast Asia countries.

|  | Net FDI inflows(USD) | ICT to total import | Infrastructure annual growth | Infrastructure. value-added | Patent | Intermediate Imports (USD) |
|---|---|---|---|---|---|---|
| mean | 4.08E+10 | 2.07E+01 | 4.96E+00 | 6.13E+11 | 1.09E+05 | 7.59E+07 |
| standard deviation | 6.90E+10 | 1.04E+01 | 4.72E+00 | 1.32E+12 | 3.19E+05 | 1.01E+08 |
| median | 9.90E+09 | 2.23E+01 | 5.01E+00 | 1.09E+11 | 6.82E+03 | 4.54E+07 |
| min | -4.95E+09 | 3.50E+00 | -1.31E+01 | 3.73E+10 | 1.15E+03 | 7.70E+06 |
| max | 2.91E+11 | 4.88E+01 | 2.38E+01 | 5.77E+12 | 1.54E+06 | 4.50E+08 |
| range | 2.96E+11 | 4.53E+01 | 3.70E+01 | 5.73E+12 | 1.54E+06 | 4.42E+08 |
| skew | 2.24 | 0.41 | -0.37 | 2.64 | 3.35 | 2.38 |
| kurtosis | 4.01 | -0.11 | 3.24 | 5.82 | 10.31 | 4.51 |

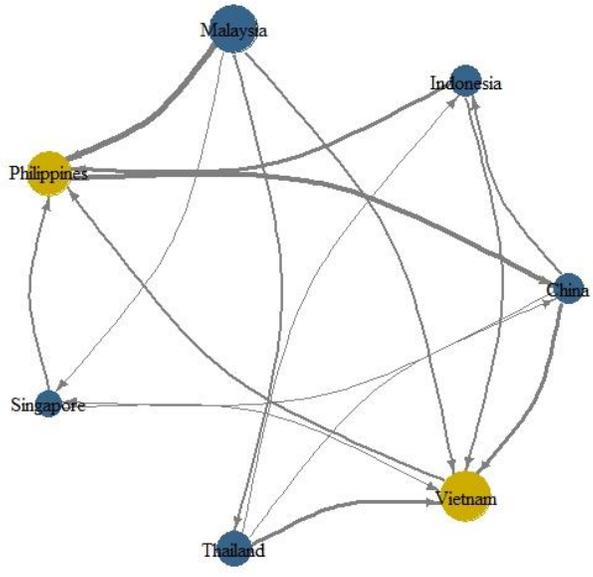

Figure A1. NPDC measure plot quantile=0.25 on immediate imports

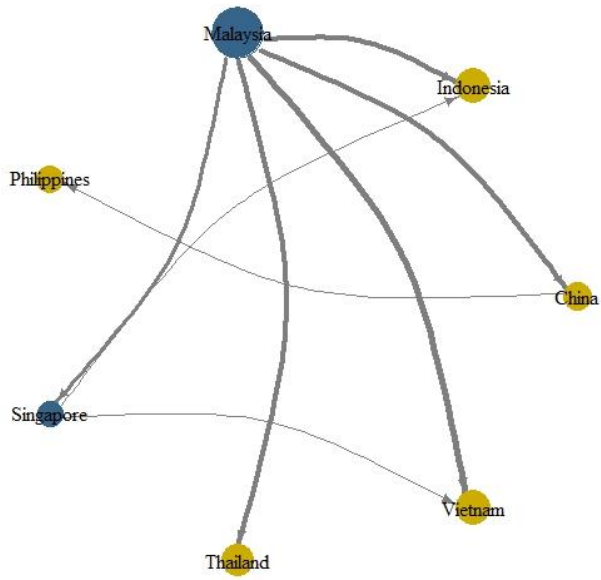

Figure A2. NPDC measure plot quantile=0.5 on immediate imports

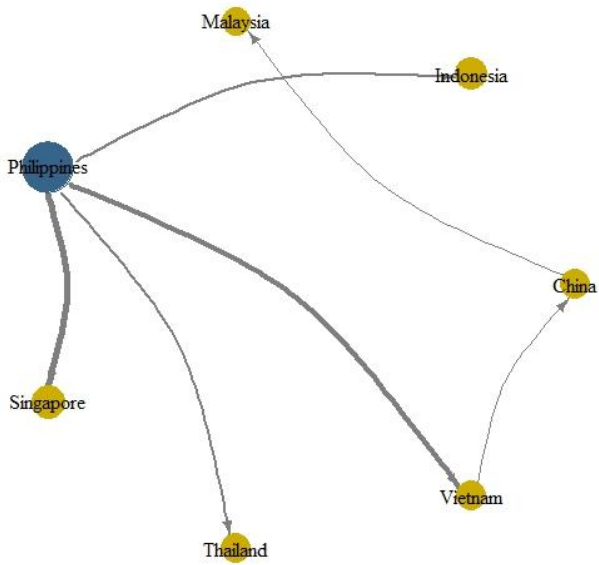

Figure A3. NPDC measure plot quantile=0.10 based on net FDI inflows

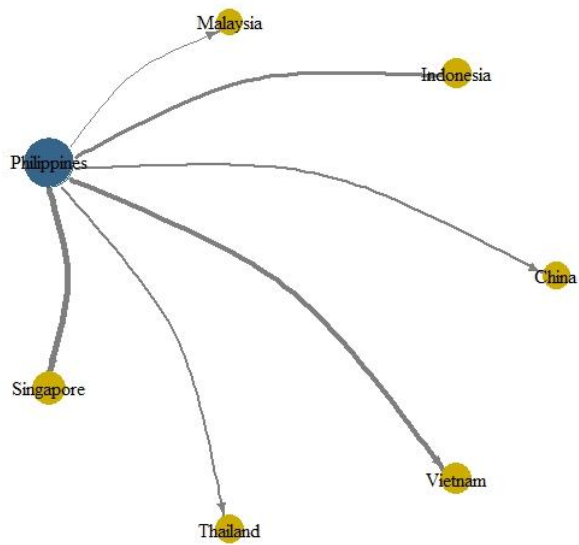

Figure A4. NPDC measure plot quantile=0.25 based on net FDI inflows

Table A2. Simulation across a sequel of quantiles on spillover contributions from other countries in terms of intermediate imports

| Quantile | China | Indonesia | Malaysia | Philippines | Singapore | Thailand | Vietnam | Total | STDEV |

| Quantile | China | Indonesia | Malaysia | Philippines | Singapore | Thailand | Vietnam | Total | STDEV |
|---|---|---|---|---|---|---|---|---|---|
| 0.05 | 79.91 | 87.48 | 93.52 | 88.59 | 88.57 | 77.21 | 92.65 | 607.93 | 6.13 |
| 0.10 | 79.91 | 87.48 | 93.52 | 88.59 | 88.57 | 77.21 | 92.65 | 607.93 | 6.13 |
| 0.15 | 77.42 | 81.5 | 80.4 | 94.88 | 84.24 | 87.85 | 90.04 | 596.33 | 6.09 |
| 0.20 | 77.42 | 81.5 | 80.4 | 94.88 | 84.24 | 87.85 | 90.04 | 596.33 | 6.09 |
| 0.25 | 77.42 | 81.5 | 80.4 | 94.88 | 84.24 | 87.85 | 90.04 | 596.33 | 6.09 |
| 0.30 | 77.42 | 81.5 | 80.4 | 94.88 | 84.24 | 87.85 | 90.04 | 596.33 | 6.09 |
| 0.35 | 73.79 | 81.95 | 81.09 | 87.16 | 86.78 | 88.71 | 91.78 | 591.26 | 6.00 |
| 0.40 | 73.79 | 81.95 | 81.09 | 87.16 | 86.78 | 88.71 | 91.78 | 591.26 | 6.00 |
| 0.45 | 74.24 | 69.82 | 72.52 | 93.72 | 91.79 | 87.45 | 95.67 | 585.21 | 11.03 |
| 0.50 | 96.56 | 97.93 | 46.52 | 93.09 | 83.8 | 96.63 | 95.41 | 609.94 | 18.53 |
| 0.55 | 99.22 | 99.31 | 39.09 | 95.42 | 86.61 | 91.74 | 69.36 | 580.75 | 21.94 |
| 0.60 | 66.74 | 99.79 | 65.49 | 87.37 | 99.83 | 99.93 | 76.51 | 595.66 | 15.56 |
| 0.65 | 88.81 | 99.44 | 65.13 | 75.1 | 96.88 | 99.1 | 86.7 | 611.16 | 13.05 |
| 0.70 | 86.3 | 75.7 | 85.95 | 99.24 | 79.27 | 81.24 | 85.79 | 593.49 | 7.52 |
| 0.75 | 86.85 | 75.59 | 85.95 | 99.24 | 79.02 | 81.4 | 85.08 | 593.13 | 7.57 |
| 0.80 | 86.85 | 75.59 | 85.95 | 99.24 | 79.02 | 81.4 | 85.08 | 593.13 | 7.57 |
| 0.85 | 86.85 | 75.59 | 85.95 | 99.24 | 79.02 | 81.4 | 85.08 | 593.13 | 7.57 |
| 0.90 | 86.85 | 75.59 | 85.95 | 99.24 | 79.02 | 81.4 | 85.08 | 593.13 | 7.57 |
| 0.95 | 86.85 | 75.59 | 85.95 | 99.24 | 79.02 | 81.4 | 85.08 | 593.13 | 7.57 |

Table A3. Simulation across a sequel of quantiles on spillover contributions to other countries in terms of intermediate imports

| Quantile | China | Indonesia | Malaysia | Philippines | Singapore | Thailand | Vietnam | Total | STDEV |
|---|---|---|---|---|---|---|---|---|---|
| 0.05 | 152.24 | 91.27 | 46.7 | 70.25 | 71.48 | 129.37 | 46.6 | 607.91 | 40.51 |
| 0.10 | 152.24 | 91.27 | 46.7 | 70.25 | 71.48 | 129.37 | 46.6 | 607.91 | 40.51 |
| 0.15 | 84.12 | 89.13 | 111.85 | 68.57 | 86.33 | 100.42 | 55.89 | 596.31 | 18.69 |
| 0.20 | 84.12 | 89.13 | 111.85 | 68.57 | 86.33 | 100.42 | 55.89 | 596.31 | 18.69 |
| 0.25 | 84.12 | 89.13 | 111.85 | 68.57 | 86.33 | 100.42 | 55.89 | 596.31 | 18.69 |
| 0.30 | 84.12 | 89.13 | 111.85 | 68.57 | 86.33 | 100.42 | 55.89 | 596.31 | 18.69 |
| 0.35 | 76.05 | 84.65 | 106.08 | 107.18 | 76.69 | 92.9 | 47.72 | 591.27 | 20.56 |
| 0.40 | 76.05 | 84.65 | 106.08 | 107.18 | 76.69 | 92.9 | 47.72 | 591.27 | 20.56 |
| 0.45 | 60.3 | 133.94 | 167.6 | 48.03 | 45.93 | 112.57 | 16.84 | 585.21 | 54.94 |
| 0.50 | 53.27 | 19.36 | 283.74 | 93.96 | 94.47 | 38.76 | 26.37 | 609.93 | 91.77 |
| 0.55 | 45.64 | 37.92 | 267.53 | 78.92 | 75.42 | 55.85 | 19.45 | 580.73 | 83.99 |
| 0.60 | 215.13 | 2.76 | 199.26 | 74.28 | 4.25 | 2.47 | 97.52 | 595.67 | 91.61 |
| 0.65 | 165.3 | 4.98 | 204.9 | 147.49 | 27.76 | 11.83 | 48.89 | 611.15 | 82.69 |
| 0.70 | 63.84 | 127.47 | 86.49 | 13.29 | 125.81 | 109.67 | 66.9 | 593.47 | 40.82 |
| 0.75 | 63.94 | 127.75 | 86.22 | 12.59 | 126.36 | 109.46 | 66.81 | 593.13 | 41.14 |
| 0.80 | 63.94 | 127.75 | 86.22 | 12.59 | 126.36 | 109.46 | 66.81 | 593.13 | 41.14 |
| 0.85 | 63.94 | 127.75 | 86.22 | 12.59 | 126.36 | 109.46 | 66.81 | 593.13 | 41.14 |
| 0.90 | 63.94 | 127.75 | 86.22 | 12.59 | 126.36 | 109.46 | 66.81 | 593.13 | 41.14 |
| 0.95 | 63.94 | 127.75 | 86.22 | 12.59 | 126.36 | 109.46 | 66.81 | 593.13 | 41.14 |

Table A4 Simulation across a sequel of quantiles on spillover contributions from other countries in terms of net FDI inflows

| Quantile | China | Indonesia | Malaysia | Philippines | Singapore | Thailand | Vietnam | Total | STDEV |
|---|---|---|---|---|---|---|---|---|---|
| 0.05 | 91.4 | 86.18 | 91.59 | 47.44 | 96.71 | 91.15 | 98.73 | 603.2 | 17.56 |
| 0.10 | 91.4 | 86.18 | 91.59 | 47.44 | 96.71 | 91.15 | 98.73 | 603.2 | 17.56 |
| 0.15 | 91.4 | 86.18 | 91.59 | 47.44 | 96.71 | 91.15 | 98.73 | 603.2 | 17.56 |
| 0.20 | 91.4 | 86.18 | 91.59 | 47.44 | 96.71 | 91.15 | 98.73 | 603.2 | 17.56 |
| 0.25 | 91.4 | 86.18 | 91.59 | 47.44 | 96.71 | 91.15 | 98.73 | 603.2 | 17.56 |
| 0.30 | 88.75 | 84.47 | 93.81 | 64.22 | 95.89 | 89.29 | 96.58 | 613.01 | 11.17 |
| 0.35 | 89.46 | 90.67 | 86.1 | 58.12 | 97.7 | 96.51 | 97.01 | 615.57 | 13.86 |
| 0.40 | 86.55 | 86.62 | 86.58 | 61.25 | 97.24 | 88.18 | 94.35 | 600.77 | 11.65 |

| 0.45 | 71.5 | 58.72 | 95.46 | 79.26 | 97.08 | 98.54 | 94.35 | 594.91 | 15.45 |
| 0.50 | 66.2 | 68.56 | 95.36 | 76.93 | 97.19 | 82.4 | 94.12 | 580.76 | 12.95 |
| 0.55 | 67.61 | 66.53 | 96.68 | 77.64 | 85.05 | 97.77 | 97.19 | 588.47 | 13.78 |
| 0.60 | 89.31 | 72.3 | 96.69 | 78.12 | 87.48 | 97.09 | 78.55 | 599.54 | 9.62 |
| 0.65 | 91.41 | 68.04 | 67.93 | 77.89 | 95.37 | 95.07 | 97.59 | 593.3 | 13.15 |
| 0.70 | 91.62 | 81.2 | 59.59 | 67.46 | 98.71 | 94.66 | 97.55 | 590.79 | 15.53 |
| 0.75 | 91.62 | 81.2 | 59.59 | 67.46 | 98.71 | 94.66 | 97.55 | 590.79 | 15.53 |
| 0.80 | 91.62 | 81.2 | 59.59 | 67.46 | 98.71 | 94.66 | 97.55 | 590.79 | 15.53 |
| 0.85 | 91.62 | 81.2 | 59.59 | 67.46 | 98.71 | 94.66 | 97.55 | 590.79 | 15.53 |
| 0.90 | 91.62 | 81.2 | 59.59 | 67.46 | 98.71 | 94.66 | 97.55 | 590.79 | 15.53 |
| 0.95 | 91.62 | 81.2 | 59.59 | 67.46 | 98.71 | 94.66 | 97.55 | 590.79 | 15.53 |

Table A5. Simulation across a sequel of quantiles on spillover contributions to other countries in terms of net FDI inflows

| Quantile | China | Indonesia | Malaysia | Philippines | Singapore | Thailand | Vietnam | Total | STDEV |
|---|---|---|---|---|---|---|---|---|---|
| 0.05 | 64.69 | 40.94 | 83.9 | 293.27 | 20.33 | 59.88 | 40.18 | 603.19 | 93.58 |
| 0.10 | 64.69 | 40.94 | 83.9 | 293.27 | 20.33 | 59.88 | 40.18 | 603.19 | 93.58 |
| 0.15 | 64.69 | 40.94 | 83.9 | 293.27 | 20.33 | 59.88 | 40.18 | 603.19 | 93.58 |
| 0.20 | 64.69 | 40.94 | 83.9 | 293.27 | 20.33 | 59.88 | 40.18 | 603.19 | 93.58 |
| 0.25 | 64.69 | 40.94 | 83.9 | 293.27 | 20.33 | 59.88 | 40.18 | 603.19 | 93.58 |
| 0.30 | 76.13 | 61.48 | 66.18 | 260.64 | 22.75 | 64.24 | 61.59 | 613.01 | 78.15 |
| 0.35 | 38.77 | 56.98 | 85.28 | 341.48 | 13.68 | 38.94 | 40.43 | 615.56 | 113.90 |
| 0.40 | 44.11 | 80.68 | 80.18 | 284.04 | 17.7 | 46.66 | 47.41 | 600.78 | 90.13 |
| 0.45 | 112.44 | 260.19 | 32.97 | 102.71 | 17.33 | 37.53 | 31.74 | 594.91 | 85.77 |
| 0.50 | 134.09 | 166.93 | 41.98 | 112.15 | 19.09 | 75 | 31.54 | 580.78 | 56.25 |
| 0.55 | 128.08 | 164.08 | 25.7 | 140.5 | 70.37 | 35.08 | 24.66 | 588.47 | 59.23 |
| 0.60 | 24.53 | 167.6 | 28.55 | 138.54 | 70.92 | 32.23 | 137.18 | 599.55 | 60.88 |
| 0.65 | 9.72 | 173.24 | 185.58 | 128.75 | 33.61 | 37.75 | 24.67 | 593.32 | 75.27 |
| 0.70 | 14.99 | 106.45 | 220.16 | 173.35 | 12.73 | 38.22 | 24.89 | 590.79 | 84.12 |
| 0.75 | 14.99 | 106.45 | 220.16 | 173.35 | 12.73 | 38.22 | 24.89 | 590.79 | 84.12 |
| 0.80 | 14.99 | 106.45 | 220.16 | 173.35 | 12.73 | 38.22 | 24.89 | 590.79 | 84.12 |
| 0.85 | 14.99 | 106.45 | 220.16 | 173.35 | 12.73 | 38.22 | 24.89 | 590.79 | 84.12 |
| 0.90 | 14.99 | 106.45 | 220.16 | 173.35 | 12.73 | 38.22 | 24.89 | 590.79 | 84.12 |
| 0.95 | 14.99 | 106.45 | 220.16 | 173.35 | 12.73 | 38.22 | 24.89 | 590.79 | 84.12 |